\documentclass{llncs}

\usepackage{makeidx}
\usepackage{amsmath, amsfonts, amssymb, mathrsfs}
\usepackage[pdftex]{graphicx}
\usepackage{url}

\begin{document}
\title{ProofFlow: Flow Diagrams for Proofs}
\titlerunning{ProofFlow}
\author{Steven A.~Kieffer}
\institute{Simon Fraser University, Burnaby BC V5A 1S6, Canada,
\email{skieffer@cecm.sfu.ca}
}

\maketitle

\begin{abstract}
We present a light formalism for proofs that encodes their
inferential structure, along with a system that transforms
these representations into flow-chart diagrams.
Such diagrams should improve the comprehensibility of proofs.
We discuss language syntax, diagram semantics,
and our goal of building a repository of diagrammatic
representations of proofs from canonical mathematical
literature. The repository will be available online in the form
of a wiki at {\tt proofflow.org}, where the flow chart drawing
software will be deployable through the wiki editor. We also consider
the possibility of a semantic tagging of the assertions in a proof, to
permit data mining.
\keywords{proofs, diagrammatic representation, flow charts,
  repositories of formalized mathematics, web presentation of
  mathematics, data mining}
\end{abstract}

%%%%%%%%%%%%%%%%%%%%%%%%%%%%%%%%%%%%%%%%%%%%%%%%%%%%%%%%%%%%%%%%%%%%%%
\section{Introduction}
\label{sec:intro}

In the spectrum of formality, from completely informal prose proofs
intended for human readers, to proof scripts checkable by automated
theorem provers or proof assistants, we propose a light level of
formalism which we shall refer to as the \emph{flow} level, in
comparison to flow charts. At this level we formally register which
assertions are said to follow from which others, thus capturing the
inferential structure of the proof, without formalizing the
assertions themselves, or demanding that the inferences be justifiable
in any particular proof system.

Proofs written in English muster their declarative statements and
assumptions into an argument using a standard battery of words and
phrases, such as \emph{so, therefore, by, then, hence, using, it
  follows that}, etc. These words provide some indication as to which
assertions are being used when others are inferred, but the
information is usually incomplete.

For example, six assertions $A_{1},A_{2},\ldots,A_{6}$ might so far have been
made in an informal proof, at which point the author might write,
``Therefore $A_{7}$.'' The author will have it in mind that $A_{7}$ follows
from $A_{2}$, $A_{5}$, and $A_{6}$, say, and will expect the reader to see this
as well, but will not make this information explicit. The word
`therefore' really just says, ``At this point you have enough to infer
$A_{7}$. Figure out how.''

At the flow level of representation of proofs, we make this
information complete and explicit.
In the language of formal natural deduction systems, a claim that an
assertion $B$ follows from assertions $A_{1},A_{2},\ldots,A_{n}$ is called a
\emph{judgment}, and may be written $A_{1},A_{2},\ldots,A_{n}\vdash B$.  In a
flow-level formalization then, while the formalizer need not add any
justification for any judgment over and above what is already present
in the proof, the judgments themselves are to be made explicit. We
show in Section \ref{sec:lang} how this is done in the ProofFlow
language.

The primary purpose of formalizing a proof in the ProofFlow system is
to improve its comprehensibility. We provide software that will take a
proof written in the ProofFlow language and produce a flow chart
representation of it.  The system is deployed through a wiki at
\url{proofflow.org}.

Many \emph{argument mapping systems} exist, and a long list can be
found at \cite{CMUargMapping}. Most of these, however, seem to cater
to the needs of real-world arguments, while none seem to address the
special demands of diagraming mathematical proof structure.
Moreover, it is suggested in \cite{KMRationale} that more
sophisticated layout is needed in such systems, and new layout
techniques are applied. The ProofFlow system could benefit from
similar techniques.

ProofFlow should aid comprehension of proofs in at least two ways: (1)
Readers are not left wondering which prior assertions should be used
in inferring a given one.  (2) The diagram gives an overview of the
proof. The reader can see at a glance how all the parts of the proof
fit together, including how many times and where in the proof each
assertion is used. In particular this should counteract the sensation
often experienced on finishing a proof, that one has confirmed each
separate inference well enough, but has lost the forest for the trees.

Though the feature has not been implemented yet, in a future version
of the system we plan to allow users to expand and collapse parts of
proofs which authors have prepared as ``further
clarifications''. Thus, a proof will be presented at an initial level
of discourse, with some inferential gaps being perhaps quite large,
while further information as to how these gaps are to be filled in may
be held in reserve. If the user requested it, this information could
be depicted by adding new nodes and directed edges to the diagram.
Expansion and collapse techniques were implemented in
\cite{KMRationale}. In \cite{KMbio} separation constraint techniques
are discussed which allow smooth alteration of graphs, so that
viewers' mental models are not disturbed.

In Sections \ref{sec:diag} and \ref{sec:lang} we discuss the semantics
of ProofFlow diagrams, and present the input language. We discuss our
goals for the \url{proofflow.org} web site in Section
\ref{sec:webSite}, and the possibility of data mining in Section
\ref{sec:DM}.

%%%%%%%%%%%%%%%%%%%%%%%%%%%%%%%%%%%%%%%%%%%%%%%%%%%%%%%%%%%%%%%%%%%%%%
\section{ProofFlow Diagrams}
\label{sec:diag}

\begin{figure}
  \includegraphics[scale=0.5]{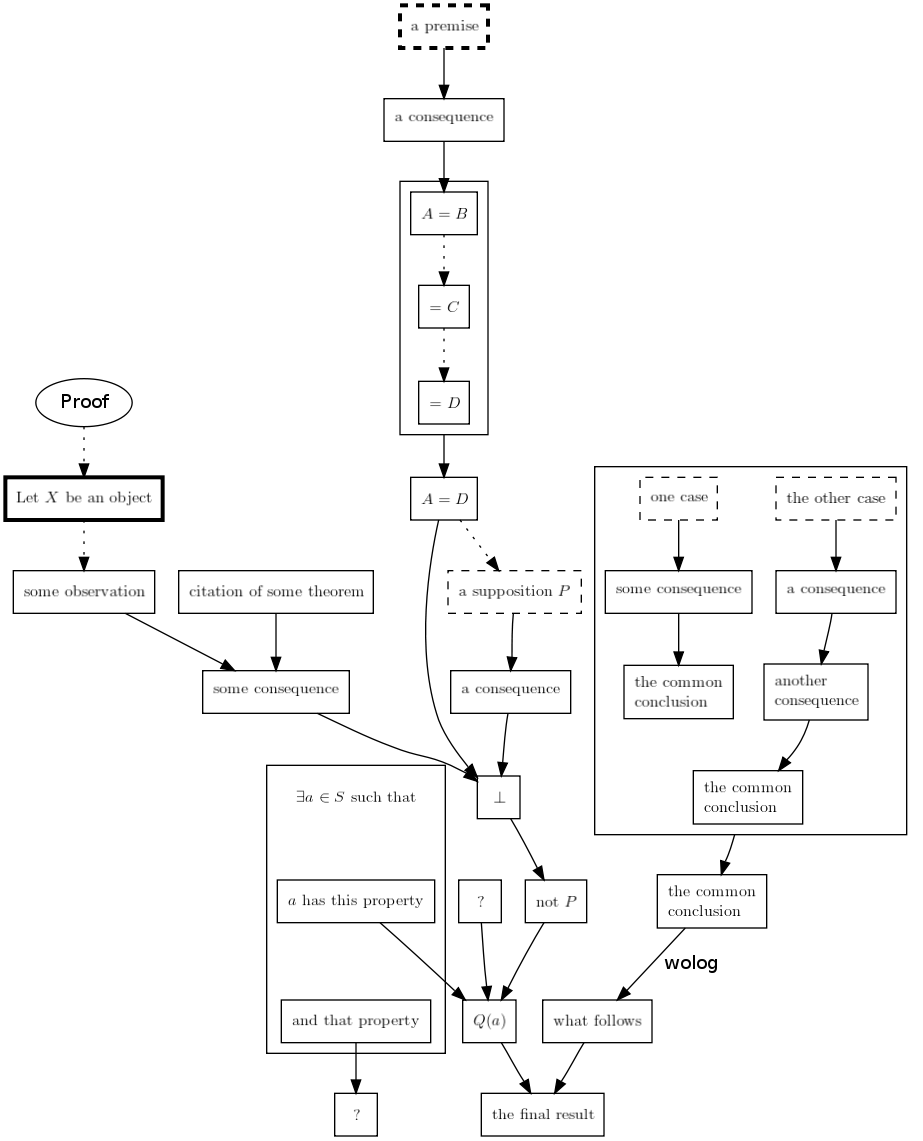}
  \caption[]{A ProofFlow diagram.}
  \label{fig:basicPFD}
\end{figure}

In a flow diagram for a proof, we put statements into boxes, and draw
arrows connecting the boxes both in order to show the inferential
structure of the proof, and in order to direct the reader through the
proof from start to finish. See for example Fig.~\ref{fig:basicPFD}.
(The input for Fig.~\ref{fig:basicPFD} is given in
Fig.~\ref{fig:input}, and will be explained in Section \ref{sec:lang}.)

\begin{figure}
  \begin{verbatim}
I01 {s Let $X$ be an object}
A02 {s some observation}
A03 {s some consequence}
C04 ( citation of some theorem )
E05 [ {s $\exists a \in S$ such that}
      {s $a$ has this property}
      {s and that property} ]
A06 {s a consequence}
A07 {s $A = B$}
A08 {s $ = C$}
A09 {s $ = D$}
A10 {s $A = D$}
A11 {s a supposition $P$}
A12 {s a consequence}
P13 {s a premise}
A14 {s not $P$}
A15 {s $Q(a)$}
A16 {s one case}
A17 {s some consequence}
A18 {l {s the common} {s conclusion} }
A19 {s the other case}
A20 {s a consequence}
A21 {l {s another} {s consequence} }
A22 {l {s the common} {s conclusion} }
A23 {l {s the common} {s conclusion} }
A24 {s what follows}
A25 {s wolog}
A26 {s the final result}

link (
    I01 go A02. So A03 by C04. Now E05. And A06 by P13.
    Now Proof
        A07 by A06 go A08 go A09
    end.
    So A10. Next suppose A11.
    Then A12, then falsum, by A10, A03.
    So A14. Then A15 by ?, E05_1.
    Now cases,
        Case A19. Then A20, so A21, so A22.
        Now Case A16. Then A17, so A18.
    end.
    So A23. Then A24 using A25. So A26 by A15.
    But ? by E05_2.
)
  \end{verbatim}
  \caption[]{The input for Fig.~\ref{fig:basicPFD}.}
  \label{fig:input}
\end{figure}

Beyond this we use a few additional graphical conventions, but overall
the diagram semantics are quite simple. The set of \emph{graphemes},
or the graphical vocabulary that we use, is very small: The boundaries
of boxes may be solid or dashed, plain or bold; the arrows between
boxes may be solid or dashed. We explain the meanings of these different
styles in the subsections below.

We use the GraphViz\footnote{\url{http://graphviz.org}} program
{\tt dot} to lay out and draw our diagrams, since they are really directed
graphs. Accordingly, we may discuss our diagrams, which we refer to as
\emph{ProofFlow} diagrams, not just in terms of boxes and arrows, but
also in terms of nodes (or vertices) and edges.

\subsection{Nodes for Assertions, Object Introductions,
    Premises, and Assumptions}
Apart from a few exceptions (to be discussed in Sections
\ref{sec:C-nodes}, \ref{sec:QF-nodes} and \ref{sec:E-nodes}),
the text appearing on each node in a ProofFlow diagram must give
either (1) a mathematical assertion, (2) the introduction of a
mathematical object, (3) a premise from the statement of the theorem
being proved, or
(4) an assumption, to be discharged somewhere in the proof. For each
of these possibilities we provide a special kind of node, whose
boundary is drawn in a special way.

Assertions are the declarative statements that make up most of a
proof. These appear on nodes with solid (i.e. not dashed), plain
(i.e. not bold) boundary, called assertion nodes, or A-nodes.

Statements defining or introducing objects into a proof will appear on
nodes with a solid, bold boundary, called introduction nodes, or
I-nodes. The bold boundary makes I-nodes easy to find, in
case the viewer wants a reminder of the definition of an object. Also
we feel that since assertions have a truth value whereas definitions
do not, it is only proper that assertions and introductions take place
in nodes that look different.

Here we introduce some technical terminology: Typically a theorem
statement will involve assumptions, and we refer to these as
\emph{premises}, whereas any additional assumptions employed during a
proof, such as for proof by cases or proof by contradiction, will be
called simply \emph{assumptions}.

In ProofFlow diagrams we write all premises and assumptions inside
boxes with dashed boundaries. In addition, the boundary is bold for a
premise, and plain for an assumption.  This makes premises and
assumptions easily distinguishable from one another, so that viewers
will not accidentally expect premises to be discharged.  We hope that
the convention is also intuitive: The dashed lines represent the
contingent nature of all premises and assumptions, whereas the bold
lines for premises represent their essential role in a theorem
statement, as opposed to the transient nature of assumptions which
will be discharged by the end of the proof.

There is a special premise node, or P-node, for premises. There is no
special node for assumptions. Instead, they are made from A-nodes, and get
their dashed boundary from the key word {\tt suppose}, as explained in
Section \ref{sec:lang}.

\subsection{Deduction}
In order to represent a judgment $A_{1},A_{2},\ldots,A_{n}\vdash B$ we draw
solid arrows, called \emph{deduction arrows}, from the $A_{i}$ to $B$.
Deduction arrows are visible in Fig.~\ref{fig:basicPFD}.

\subsection{Scoping and Flow}
When we read a prose proof we always know what objects have been
introduced since they are defined before they are used, and we always
know what assumptions are in force since they are stated before they
are used. Questions as to where in a proof objects or assumptions are
in effect are questions of \emph{scope}.

In a flow diagram for a proof we stand to find ourselves confused on
matters of scope, since we throw away much of the linear order of the
prose, and we are liable to view the nodes in the graph in the wrong
order.

In order to avoid this problem we employ what we call
\emph{flow arrows} in a ProofFlow diagram. A flow arrow from node $A$
to node $B$ is drawn dashed rather than solid, and rather than saying
that $A$ is used in deducing $B$ it simply says that after you have
considered $A$ the author of the proof wants you to consider $B$ next.
See Fig.~\ref{fig:basicPFD} for examples.

So that the proof has a starting point, we automatically include
a single oval-shaped node reading ``Proof,'' which points with a flow
arrow toward the first node declared in the proof. (See Section
\ref{sec:lang} on node declarations.)

In this way we ensure that the viewer is directed through a ProofFlow
diagram in an appropriate order, so that he or she will not encounter
a statement about an object $X$ without already having seen the
definition of $X$, and so that it will be clear which
assumptions are in force at any given point.

While in completely formal proofs, such as Fitch-style proofs, both
the beginning \emph{and the end} of the scope of any assumption is
made clear by means of clearly delimited subproofs, in general we get
away without taking such care in prose proofs. The linear order of the
discussion seems to suffice. Flow arrows therefore should also suffice
in ProofFlow diagrams. If not, formalizers may choose to put
certain deduction chains into boxed subproofs.
(See Section \ref{sec:subproofs}.)

Formalizers and viewers of ProofFlow diagrams should follow these
rules regarding the use of flow and deduction arrows:
\begin{enumerate}

\item There should not be both a flow and a deduction arrow going from
  a node $A$ to a node $B$; in this case only the deduction arrow
  should be used; the viewer will understand that it is to be
  followed.

\item A single node $A$ might have a flow arrow to a node $B$ and
  deduction arrows to one or more other nodes $C_{1},\ldots,C_{n}$. In this
  case the viewer should follow the flow arrow; the deduction arrows
  will naturally come into consideration later in the proof.

\item There should never be more than one flow arrow leaving a single
  node.

\item In general, the viewer should never be left wondering where to
  go next.
\end{enumerate}

\subsection{Citation Nodes}
\label{sec:C-nodes}
When a theorem or lemma is cited in support of a claim, we put the
name of the result in a special citation node, or C-node, and draw a
deduction arrow from the C-node to the claim. The \url{proofflow.org}
web site is still largely under construction, but when more
infrastructure is in place then C-nodes will provide clickable links to
the results that they name.

\subsection{Question and Falsum Nodes}
\label{sec:QF-nodes}
When the contradiction is reached in a proof by contradiction, this
should be shown in a ProofFlow diagram by deducing an A-node
displaying only the falsum symbol, $\bot$. We provide a convenient way
to generate such a node, which we discuss in Section \ref{sec:lang}.

A formalizer may be unsure about certain parts of a proof, and in
this case \emph{question nodes} should be used. A question node is an
A-node displaying only a question mark, and again we provide a
convenient way to generate such a node, to be discussed in Section
\ref{sec:lang}.

Specifically, if the formalizer feels that a (possibly empty) set of
nodes $\{A_{1},\ldots,A_{n}\}$ are used in inferring node $B$, but that
something more is also needed, though he or she is not sure what, then
in addition to a deduction arrow from each $A_{i}$ to $B$, there should
also be a deduction arrow from a question node to $B$.

On the other hand, if the formalizer does not know where an assertion
$C$ is used in a proof, but feels that it should be used
somewhere, then there should be a deduction arrow from $C$ to a
question node. Care should be taken to maintain flow in this case; in
particular, it may be necessary to draw a flow arrow leaving the
question node. See Fig.~\ref{fig:basicPFD} for demonstration of both
uses of question nodes.

\subsection{Existence and Introduction}
\label{sec:E-nodes}
In a highly formal proof, it is one
thing to say that an object of a certain kind exists, and it is
another thing to introduce such an object. These form two separate
steps in the proof. In informal proofs, however, these steps
usually occur simultaneously, with the existence being stated
explicitly, and the introduction of such an object being implicit.

For example, the author of a proof in group theory might say that
there exists a group homomorphism $\varphi$ having certain properties,
and in the next breath proceed to talk about $\varphi$. As the reader of
the proof we understand that $\varphi$ now refers to an arbitrary object
of the kind just stated to exist.

In order to accurately model informal proofs, we too combine these
steps in ProofFlow diagrams.  After drawing a node that states the
existence of an object, the formalizer may then proceed immediately to
draw nodes that discuss such an object.

It is important however that later on in the proof we have a way to
cite not just the existential statement itself, but also the
individual statements of the properties of the existing object. For
this reason we provide a special existence node, or E-node, with
which to make the existence statement. (Its syntax is described in
Section \ref{sec:lang}.)

Let us consider for example the following three steps from the proof
of Theorem 90 in \cite{Zahlbericht}:
\begin{enumerate}
\item $\exists$ an integer $a\in\mathbb{Z}$ such that $\psi(a)\neq0$.
\item Define $B^*=\frac{\psi(a)}{a+\theta}$.
\item $A=\left(B^*\right)^{1-S}$.
\end{enumerate}
In line 1 we state the existence of an integer $a$ with a certain
property, in line 2 we then go on to use such an integer $a$ to
construct a number $B^*$, and in line 3 we assert an equation
involving $B^*$.

When we infer line 3, we rely on the fact that $\psi(a)\neq0$,
\emph{not} on the \emph{existence} of such an integer $a$. Therefore
in a ProofFlow diagram we need to be able to draw an arrow not from a
box around the existential statement, but from another box, around the
statement that $\psi(a)\neq0$. E-nodes provide for this in ProofFlow
diagrams. See example at bottom of Fig.~\ref{fig:basicPFD}.

\subsection{Subproofs}
\label{sec:subproofs}
Sometimes in an informal proof we make a claim, and then devote a
special subproof to the support of this claim. Such a subproof will be
depicted in a ProofFlow diagram by surrounding all the nodes of the
subproof within a box.

Deduction and flow arrows may be drawn from the surrounding box of a
subproof to other nodes in the diagram. The box is also
``transparent'' in that it is still possible to draw arrows connecting
the nodes inside to nodes outside the subproof. It is not possible to
draw a deduction arrow \emph{to} a subproof, since a subproof is not
the sort of thing which, in its entirety, is to be supported by a
reason! A flow arrow can however be drawn to a subproof.

One particular type of subproof is proof by cases: We consider a
series of assumptions $A_{1},\ldots,A_{n}$ in order, and show that under each
assumption (without any of the others) we can reach one and the same
conclusion $C$. While we entertain the possibility of using special
graphemes for proof by cases in the future, at present we simply
depict each of the assumptions $A_{i}$ with a dashed boundary, as with
any other assumption, and collect each of the $n$ deductions from $A_{i}$
to $C$ together inside a subproof box. See key words {\tt cases} and
{\tt case} in Section \ref{sec:lang} for further details.

\subsection{Relation Chains}
Relation chains are a common feature in informal proofs. These may be
chains of equations, inequalities, subset relations, or other infix
relations. Such displays might have the following form:
\begin{align*}T_{1}&=T_{2}\\
&=T_{3}\\
&\vdots\\
&=T_{n}.\end{align*}

Since each relation in the chain represents a separate assertion which
may need to be justified or may be used in later justifications, there
should be a separate box around each relation, in a ProofFlow
diagram.

Therefore we make the following graphical convention for relation
chains: The first relation (e.g. $T_{1}=T_{2}$) should go in a node $A_{2}$,
and subsequent relations (e.g. $=T_{3}$, ..., $=T_{n}$) should be written,
starting with the relation symbol, in nodes $A_{3},\ldots,A_{n}$.  Flow
arrows should be used to link $A_{2}$ to $A_{3}$, $A_{3}$ to $A_{4}$, and so forth
up to $A_{n}$.  See again Fig.~\ref{fig:basicPFD} for an example.

\subsection{Edge Labels}
A label may be put along a deduction arrow in order to give some
indication of the method of inference that is being used.  At present
we support only text, not graphical, edge labels.

Two important examples will be the labels ``by induction,'' and
``wolog'' or ``without loss of generality''.  Formally speaking,
``wolog'' is really a way of saying, ``We're going to do proof by
cases, but we're actually only going to discuss one of the cases,
since all the others can be reduced to it.'' In a ProofFlow diagram,
the assumption $A$ introduced ``without loss of generality'' should be
led to by an arrow labeled ``wolog''.

%%%%%%%%%%%%%%%%%%%%%%%%%%%%%%%%%%%%%%%%%%%%%%%%%%%%%%%%%%%%%%%%%%%%%%
\section{The Language}
\label{sec:lang}
The ProofFlow language is very simple. In an input file, the user
first declares all nodes, giving them names and giving the text that
should appear in them. The user then writes a paragraph called a
\emph{proof script} that links the nodes together by edges, labels
edges if desired, makes nodes into assumptions as appropriate, and can
put nodes into subproofs. See Fig.~\ref{fig:input} for the
input file that produced Fig.~\ref{fig:basicPFD}.

Each node in a ProofFlow diagram gets a name, which may use only
capital and lowercase letters, and digits. The first character must be
one of the capital letters in the set {\tt [ACEIP]}, and this
indicates what \emph{type} of node it is. We review the different
types of node below.

Recommended practice is to simply number the nodes in order of their
use in the proof, so that the name of each node consists of an initial
letter indicating its type, followed by a numeral. For example:
{\tt I00, A01, A02, A03, C04, E05,} and so forth.

\subsection{Text Objects}
The text displayed on nodes is typeset using \LaTeX{}.  Since a node is
a small rectangular box, we will often want the text to be laid out not
as it would be on a page, but rather in short lines arranged in a
left- center- or right-aligned column. The user may achieve this
through standard \LaTeX{} commands, but we also provide convenient
shortcut syntax for this purpose.

Text is placed inside a \emph{text object}, which is delimited by
braces {\tt \{\}} and which must begin either with a sequence of the
letters {\tt [lcr]} to indicate that a table environment with left-
center- and right-aligned columns is desired, or else with the letter
{\tt s} to indicate that no such formatting is desired. Text objects
may be nested to as great a depth as \LaTeX{} will accept.

For example, the text object
\begin{verbatim}
     {l
         {s $K$ a number field,}
         {s $L$ its normalization.}
     }
\end{verbatim}
produces two left-aligned lines of text:
\[\begin{array}{l}K\mbox{\:a number field,}\\
L\mbox{\:its normalization.}\end{array}\]

\subsection{Node Types}

\subsubsection{Assertion Nodes or A-nodes.}
The name of an assertion node must begin with an {\tt A}.
To declare an assertion node the user simply types the name of the
node, followed by a text object giving the text that should appear on
the node. For example:
\begin{verbatim}
     A01 {s $a \in K$}
\end{verbatim}

\subsubsection{Citation Nodes or C-nodes.}
A citation node is used in order to cite a separate theorem or lemma.
The name of a citation node must begin with a {\tt C}.
It takes a single argument given in parentheses {\tt ()}, which gives
the name of the result being cited, and within the context of the wiki
at {\tt proofflow.org} will also provide a hyperlink to a page devoted
to that result. (As of the time of this writing, the hyperlink feature
is not yet implemented.) For example:
\begin{verbatim}
     C02 ( Zahlbericht_Thm_148 )
\end{verbatim}

\subsubsection{Existence Nodes or E-nodes.}
Recall from Section \ref{sec:E-nodes} that existence nodes
achieve the special purpose of both stating the existence of an object
or objects, and at the same time introducing those objects.  An existence node
must have a name beginning with an {\tt E}.  It takes a
square-bracket-delimited sequence of text objects as its argument. For
example:
\begin{verbatim}
  E03 [
    {s $\exists$ an integer $a \in \mathbb{Z}$ such that}
    {s $\psi(a) \neq 0$.}
  ]
\end{verbatim}

In general, if $t_{0},t_{1},\ldots,t_{n}$ are the text objects passed to an
existence node named {\tt E{\it name}}, then $t_{0}$ should name a
mathematical object and state its existence, while $t_{1},\ldots,t_{n}$
should state the properties of the object.  In the above example, the
two text objects
\begin{center}
\begin{tabular}{rcl}
$t_{0}$ & : & $\exists$ an integer $a\in\mathbb{Z}$ such that \\
$t_{1}$ & : & $\psi(a)\neq0$
\end{tabular}
\end{center}
follow this rule. In order to refer to the
individual text objects
$t_{0},t_{1},\ldots,t_{n}$
later, they are automatically assigned the
names {\tt E{\it name}\_0}, {\tt E{\it name}\_1}, ...,
{\tt E{\it name}\_$n$}. In our example, the assertion $\psi(a)\neq0$ will
appear in a boxed node that can be referenced as {\tt E03\_1}.

Note however that whereas each of the two text objects $t_{0}$, $t_{1}$ in
our example will have a box around it so that it can play a role in
the proof, we cannot meanwhile cite the substatement of $t_{0}$ which
says ``$a\in\mathbb{Z}$.''  This might be desirable, as a way of saying,
``Here we are using the fact that $a$ is an integer.'' While at
present there is no way to do this in the ProofFlow syntax, we hope to
provide this feature in the future.

\subsubsection{Introduction Nodes or I-nodes.}
An introduction node has a name beginning with {\tt I}, and introduces
new objects into the proof. It takes a single text object as argument.
For example:
\begin{verbatim}
     I04 {s Let $L$ be the normalization of $K$. }
\end{verbatim}

\subsubsection{Premise Nodes or P-nodes.}
The name of a premise node begins with the letter {\tt P}. It recalls
one of the premises made in the statement of the theorem being proved.
It takes a single text object as argument.
For example:
\begin{verbatim}
     P05 {s $N(a) = 1$}
\end{verbatim}

\subsection{The Linking Language}
Once all nodes have been declared it is time to link them together
into a proof. To achieve this the user writes a paragraph in the
ProofFlow linking language, which we call a \emph{proof script}, and
passes it to the {\tt link} function, as in Fig.~2. The language uses
only the names of the declared nodes together with the following key
words (words written on the same line are synonyms):
\begin{verbatim}
     so, then
     by
     go, next
     now, and, but
     suppose, case
     proof, cases
     end
     using
     ?
     falsum
\end{verbatim}
White space is ignored, the key words are \emph{not} case-sensitive,
and the three punctuation symbols {\tt [;,.]} are also ignored. Thus
the paragraph may be organized into ``sentences'' if desired, although
this is not necessary. See the example at the bottom of
Fig.~\ref{fig:input}.

\subsubsection{The Keywords.} \hfill

\medskip \noindent {\tt by}:
Key word {\tt by} introduces supporting reasons. Thus,
\begin{verbatim}
        A0 by A1 A2 A3 ...
\end{verbatim}
will cause deduction arrows to be drawn to node {\tt A0} from each of
the nodes
{\tt A1, A2, A3, ...}.

In general the head of the deduction arrows will be the most recent
node named. Thus,
\begin{verbatim}
        A0 by A1 by A2
\end{verbatim}
will yield two deduction arrows: one from {\tt A1} to {\tt A0}, and
one from {\tt A2} to {\tt A1}.

\medskip
\noindent {\tt so, then}: Key word {\tt so} (and synonym {\tt then})
introduces logical consequences. Thus,
\begin{verbatim}
        A0 so A1 A2 A3 ...
\end{verbatim}
will cause deduction arrows to be drawn from node {\tt A0} to nodes
{\tt A1, A2, A3, ...}.

In general the tail of the deduction arrows will be the most recent
node that has been \emph{deduced} (not named as a supporting
reason). Thus,
\begin{verbatim}
        A0 by A1 so A2
\end{verbatim}
will yield two deduction arrows: one from {\tt A1} to {\tt A0}, and
one from {\tt A0} to {\tt A2}.

\medskip \noindent {\tt go, next}:
Key word {\tt go} (and synonym {\tt next}) introduces a flow arrow.
Thus,
\begin{verbatim}
        A0 go A1
\end{verbatim}
draws a flow arrow from node {\tt A0} to node {\tt A1}.
As with {\tt so} and {\tt then}, the tail of the flow arrow will be
the most recent node that has been deduced.

\medskip \noindent {\tt now, and, but}: Key word {\tt now} and its
synonyms allow the user to introduce a node that is not connected to
any of those already named. Thus, if we have just shown that node
{\tt A0} implies node {\tt A1}, say, and want to now start a separate
chain of reasoning, in which node {\tt A2} will imply node {\tt A3},
we may write
\begin{verbatim}
        now A2 so A3
\end{verbatim}

\medskip \noindent {\tt suppose}: If the key word {\tt suppose} is
placed before a node {\tt A1} then node {\tt A1} will be taken as an
assumption, and will be given a dashed boundary.

\medskip \noindent {\tt proof} and {\tt end}: The key word
{\tt proof} starts a subproof, and the key word {\tt end} ends
it. Everything inside the subproof will be drawn inside a box in the
ProofFlow diagram. Syntactically, the entire subproof can be used
precisely as any node name is used, except that a subproof cannot come
after the {\tt so} and {\tt then} key words. (Thus, a subproof can be
a reason, but cannot be a consequence.)

\medskip \noindent {\tt cases} and {\tt case}:
The key word {\tt cases} is meant to initiate a special kind of
subproof, namely, a proof by cases. Within such a subproof the key
word {\tt case} indicates the start of a new case. The entire subproof
is terminated by the key word {\tt end}.

In future versions of the system we may use more specialized graphical
conventions to represent proof by cases, but for now {\tt cases} is
just a synonym for {\tt proof}, and {\tt case} is just a synonym for
{\tt suppose}. The user should not forget to put {\tt now} (or a
synonym) before each {\tt case} after the first.

\medskip \noindent {\tt using}:
After a deduction arrow is introduced by either the {\tt so} or
{\tt by} key words or their synonyms, the key word {\tt using} will
put a label on the arrow. The label for the arrow should have been
declared in an assertion node, using plain text in the text object for
the node, since the text will \emph{not} be typeset by \LaTeX{}.

\medskip \noindent {\tt falsum} and {\tt ?}: Anywhere in the
proof script where a node name may appear, the user may write
{\tt falsum} or {\tt ?}, which will automatically generate a node
featuring only a typeset `$\bot$' or `?', respectively. As was
discussed in Section \ref{sec:QF-nodes}, the former is for use in proof by
contradiction, and the latter in case the author of the proof script
is unsure how an inference is to be made.

If it is clear, for example, that {\tt A1} and {\tt A2} were used in inferring
{\tt A0}, but the formalizer thinks that something more still is needed,
then a question node should be used, as in:
\begin{verbatim}
        A0 by A1 A2 ?
\end{verbatim}

On the other hand, if an assertion {\tt A3} has been made in the proof, but
the formalizer does not see where {\tt A3} is used, then a question
node should be placed as the consequence of {\tt A3}, as in:
\begin{verbatim}
        A3 so ?
\end{verbatim}

\subsubsection{Putting it Together.}

A complete proof script might read something like the example at the
bottom of Fig.~\ref{fig:input}.

%%%%%%%%%%%%%%%%%%%%%%%%%%%%%%%%%%%%%%%%%%%%%%%%%%%%%%%%%%%%%%%%%%%%%%

\section{Conclusions and Outlook}

\subsection{Building a Catalog of Proofs}
\label{sec:webSite}

The ProofFlow system is available in a wiki at {\tt proofflow.org}, where we
aim to build a catalog of proofs displayed in flow diagrams. The
site is built on the Mediawiki wiki
engine,\footnote{\url{http://www.mediawiki.org/wiki/MediaWiki}}
and the ProofFlow software is deployed via an extension
to the wiki editor: ProofFlow input is to be placed inside
a pair of {\tt <proofflow>}, {\tt </proofflow>} tags.

It is hoped that the ProofFlow website will be used as an aid in
exploring dense mathematical literature, in particular classic or
canonical works. We have begun with Hilbert's \emph{Zahlbericht}
\cite{Zahlbericht}.

Ultimately we hope that a large corpus of proofs from classic
mathematical literature will be formalized, and that the web site will
facilitate studies in the history of ideas in proofs.

\subsection{Data Mining}
\label{sec:DM}

Besides facilitating aids to the comprehensibility of a proof such as
flow diagrams, the flow level of formalization also provides
interesting opportunities for data mining.

By formalizing the inferential structure of a proof
as it is presented to human readers, we provide
ourselves with a data set in which we can hope to discover patterns of
reasoning at a tactical level on which mathematicians are accustomed
to thinking.

We might hope that mining for patterns at this level could reveal
generalizations about writing proofs comparable to those that are
often made about a tactical but enormously complex game like chess,
where advice takes the form of statements such as ``control the
center,'' ``activate your minor pieces early,'' etc. The prospects
might be improved here if we were to concentrate on one area of
mathematics at a time, say, algebraic number theory.
Perhaps we would learn generalizations about the kinds of objects that
are often constructed in proofs under certain circumstances, for example.

The possibilities for data mining could be improved if some
semantic markup was added to the assertions stated on the nodes of the
proof. Of course when we ask users to add semantic markup we ask them
to do more work than is necessary in order to see a nice diagram, and
it is therefore reasonable to consider lightweight or easy markup
systems that might be less off-putting than full formalization.

One easy technique which could improve searchability of the
database would be the mere indication of the \emph{type} of each
object named in the proof. Thus, if $K$ is a number field, the user
should say so; if $a$ is an element of $K$, the user should say
so. Even if we asked nothing more than this, searches through the text
written on the nodes would be much more meaningful, since we would
know what kinds of objects were being discussed.
A markup scheme like this could be enforced by requiring that type
arguments be passed to I-nodes.

We hope to include such a light-weight markup system, and even a
heavy-weight alternative, in the ProofFlow system in the future.

\subsection{Questions}
\label{sec:CandO}

The diagram semantics employed in ProofFlow diagrams remain
experimental. If enough proofs are formalized, and enough users
comment on the comprehensibility of the diagrams, we will learn
whether our graphical conventions work well or not.

In particular we hope to answer the following questions: Will proper
use of flow arrows always make it clear which assumption is being
negated in a proof by contradiction? Will proof by cases be clear?
While it is expected that in most real-world proofs the logic seldom
gets terribly complex, will it be possible in exceptionally complex
cases to use subproofs to organize a ProofFlow diagram so that the
logic remains clear and unambiguous?

The highest priority in improving the system at present is to allow
authors to include optional clarifications for inferential gaps, and
viewers to show and hide the corresponding subgraphs, as discussed in
Section \ref{sec:intro}. Ideally, graph transformations might be
performed as animations, using smooth graph redrawing techniques such
as have been discussed in work such as \cite{KMbio}, \cite{Animated}.

Overall the hope for \url{proofflow.org} is that it will provide users
with a new way to read and understand proofs, and a collaborative
environment in which the logic of proofs can be worked out, discussed,
and elaborated upon. Perhaps it can offer a new gateway into classic
mathematical literature. In the future, depending on how much effort
we put into the addition of semantic markup, we may mine the data in
attempt to learn interesting tactical generalizations about proofs
belonging to various areas of mathematics.

%%%%%%%%%%%%%%%%%%%%%%%%%%%%%%%%%%%%%%%%%%%%%%%%%%%%%%%%%%%%%%%%%%%%%%

\end{document}